\documentstyle[sprocl,epsfig]{article}

\def\NPB{{\em Nucl. Phys.} B}
\def\PLB{{\em Phys. Lett.}  B}

\def\PRD{{\em Phys. Rev.} D}
\def\EPJ{{\em Eur. Phys. J.} C}

\def\be{\begin{equation}}
\def\ee{\end{equation}}
\def\bea{\begin{eqnarray}}
\def\eea{\end{eqnarray}}

\newcommand\ch{{\tilde{\chi}}}
\newcommand\ti{\tilde}
\renewcommand\d{\delta}

\begin{document}

\title{Higgs boson decays into charginos and neutralinos\\
including full one-loop corrections\\}

\author{H.~Eberl}

\address{Institut f\"ur Hochenergiephysik der \"Osterreichischen
 Akademie der Wissenschaften, A-1050 Vienna, Austria}


\maketitle\abstracts{
We present the one-loop corrected decay widths of a neutral
(charged) Higgs boson into a chargino or neutralino pair (chargino and neutralino)
in the CP-conserving MSSM. The chargino and neutralino parameters 
are renormalized
in the on-shell scheme and it is shown that the correction due to the chargino/neutralino 
mass matrix renormalization is comparable to the conventional vertex correction. 
The full corrections are
typically of the order of 10\,\%. 
}

\section{Introduction}
The Minimal Supersymmetric Standard Model (MSSM) 
looks still the most attractive extension of the Standard Model.  
Taking the input parameters real in the MSSM, there are 
two CP-even neutral bosons ($h^0$, $H^0$), one CP-odd boson $A^0$, 
and two charged bosons $H^{\pm}$. For the verification of the MSSM, 
detection and precision studies of these Higgs bosons are 
necessary. If $(H^0, A^0, H^\pm)$ are heavy, many decay modes are possible, 
especially the decay channels into the SUSY particles squarks, sleptons, charginos
and neutralinos can be open. In this work, we study the decays
\begin{eqnarray}
(H^0, A^0) & \to & \ch^+_i + \ch^-_j\,, \,  \ch^0_m + \ch^0_n\,;\\
H^+ & \to & \ch^+_i + \ch^0_m\, .
\end{eqnarray} 
These decays originate from Higgs-gaugino-higgsino interaction.
They are therefore sensitive to the $\ch$-mixing 
($\ch$ denotes chargino or neutralino) and of particular
interest. 
We calculate the full one-loop corrected decay widths including soft and hard 
photon bremsstrahlung. In addition to the conventional wave-function and vertex 
corrections, the corrections due to the $\ch$-mass matrix 
renormalization
\cite{chmasscorr1,chmasscorr2} has to be included. In this work we apply
the procedure of \cite{chmasscorr1}. The full one-loop corrected widths for 
neutral Higgs boson 
decays into charginos are already published in \cite{Higgs->charginos}.
The (s)fermionic correction to the neutral Higgs boson decays
into charginos are given in \cite{H0->neutralinos} and
leading Yukawa (s)top and (s)bottom corrections to the decays into
a chargino/neutralino pair in \cite{zhang}. 
 
\section{One-loop decay width}

The tree-level decay width of $H_k \to \ti \chi_i \,\ti \chi_j$
with $H_k = \{ h^0, H^0, A^0, H^+ \}$, see e.g. \cite{tree2},
is given by 
\begin{equation}
\Gamma^{\rm tree} =
 \frac{g^2}{16 \pi m_{H_k}^3}\, \frac{\kappa}{1 + \delta_{ij}}\, \left(
X\, (F^2_{ijk} + F^2_{jik}) - 4 \eta_k m_{\ti \chi_i} m_{\ti \chi_j} F_{ijk} F_{jik}\right)\,,
\end{equation}
with $X = (m_{H_k}^2 - m_{\ti \chi_i}^2 - m_{\ti \chi_j}^2)$,
$\kappa = (X^2  - 4  m_{\ti \chi_i}^2  m_{\ti \chi_j}^2)^{1/2}$, 
$\eta_{1,2} = 1$ for the CP even states
$h^0$ and $H^0$, and $H^+$, and $\eta_3 = -1$ for the CP odd state $A^0$, 
$\delta_{ij}$ is only necessary for 
the decays into two neutralinos.
The term $g F_{ijk}$ denote the chargino/neutralino-Higgs boson 
, e.g.
the chargino-neutral Higgs boson coupling is defined in \cite{Higgs->charginos}.

\noindent The corrected width $\Gamma^{\rm corr} = \Gamma^{\rm tree} + \Delta \Gamma$ with 
the ultraviolet and infrared convergent one-loop contribution is
\begin{eqnarray}
\hspace*{-3mm}
\Delta \Gamma 
 &=& \frac{g^2}{8 \pi\, m^{3}_{H_{k}} }\,\frac{\kappa}{1 + \delta_{ij}} \,
\left[ X\, {\rm Re} (F_{ijk} \Delta F_{ijk} + F_{jik} \Delta F_{jik} ) 
\right. 
\nonumber \\ 
&& \left. 
 - 2 \eta_k  m_{\ti \chi_i}  m_{\ti \chi_j} {\rm Re}
(F_{ijk} \Delta F_{jik} + F_{jik} \Delta F_{ijk} )  \right] 
+ \Gamma(H_{k} \to \ch_i\, \ch_j\, \gamma ) \, . 
\label{eq:gammacorr}
\end{eqnarray}
The one-loop correction to the coupling $F_{ijk}$ is 
\begin{equation}
  \label{eq:Fren}
F^{\rm corr.}_{ijk} = F_{ijk} + \Delta F_{ijk}
= F_{ijk}+ \d F^{(v)}_{ijk} + \d F^{(w)}_{ijk} +
 \d F^{(c)}_{ijk} \, , 
\end{equation}
where $\d F^{(v)}_{ijk}$, $\d F^{(w)}_{ijk}$, and $\d F^{(c)}_{ijk}$ 
are the vertex correction, the wave-function correction, and the 
counter terms for the parameters $g,g', U, V,$ 
$Z, \alpha, \tan\beta$
given in $g F_{ijk}$, respectively. For $A^0$ the term $\d F^{(w)}$ includes
$A^0$-$Z^0$ and $A^0$-$G^0$, and for $H^+$ the $H^+$-$W^+$ and $H^+$-$G^+$ mixing.
In the following "naive correction" 
denotes corrections due to $\Delta \Gamma$. 
The shift due to the $\ch$-mass matrix renormalization is already
included in $\Gamma^{\rm tree}$.
Further details see \cite{chmasscorr1,Higgs->charginos}.  
All one-loop amplitudes were calculated using the packages by T.~Hahn \cite{FeynArts}.
The (s)fermion loops and all one-loop self-energies were calculated also 
analytically. We also checked the consistency within the approach of \cite{zhang}.

\section{Numerical results}

The SUSY parameter set of SPS1a of the Snowmass Points and Slopes is chosen as 
reference point \cite{SPS1a}. The detailed input parameters can be read in \cite{Higgs->charginos}.
For the renormalization of the SU(2) gauge coupling $g$ we used $\alpha(m_Z)$ as input. 
We also worked out the scheme with the
Fermi coupling constant $G_F$ as input. Numerical comparison showed that the difference between these
two schemes is mainly a higher order effect. For $\tan\beta$ we used the on-shell value as input parameter.
Using the $\overline{\rm DR}$ value as input, we checked that for these processes the numerical 
difference is small.  

\begin{figure}[h!]
 \begin{center}
\hspace{2mm}
\mbox{
\mbox{\resizebox{55mm}{!}{\includegraphics{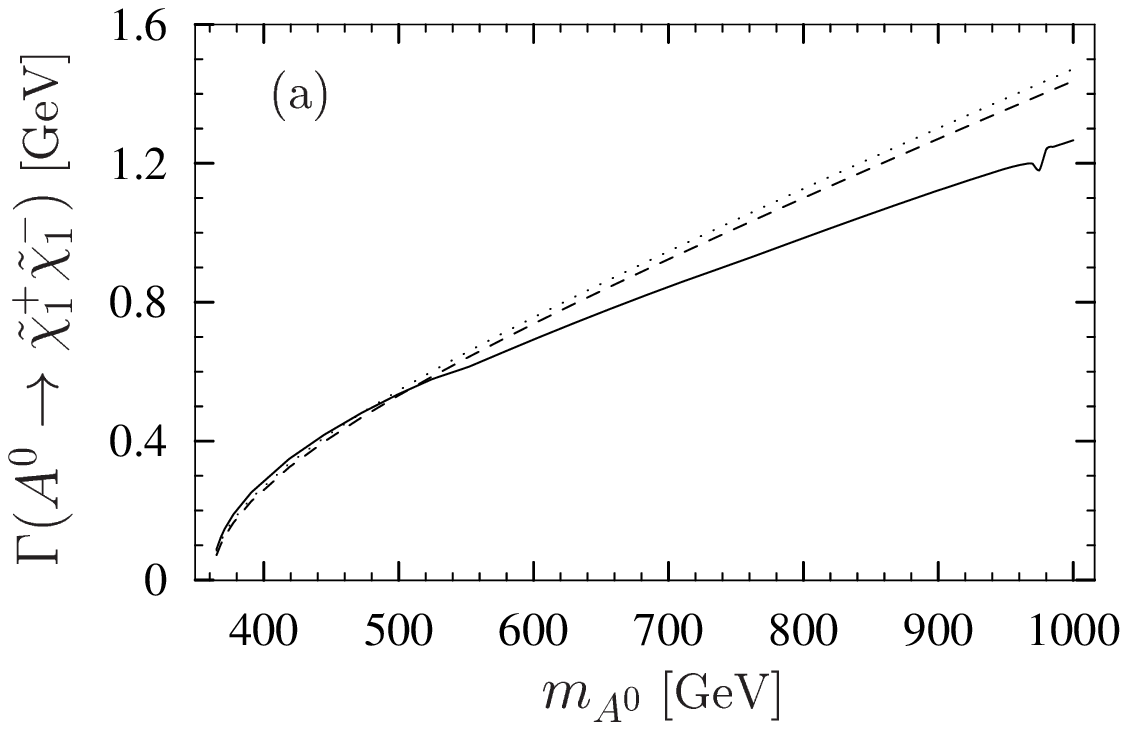}}}
\raisebox{-0.7mm}{\mbox{\resizebox{60.3mm}{!}{\includegraphics{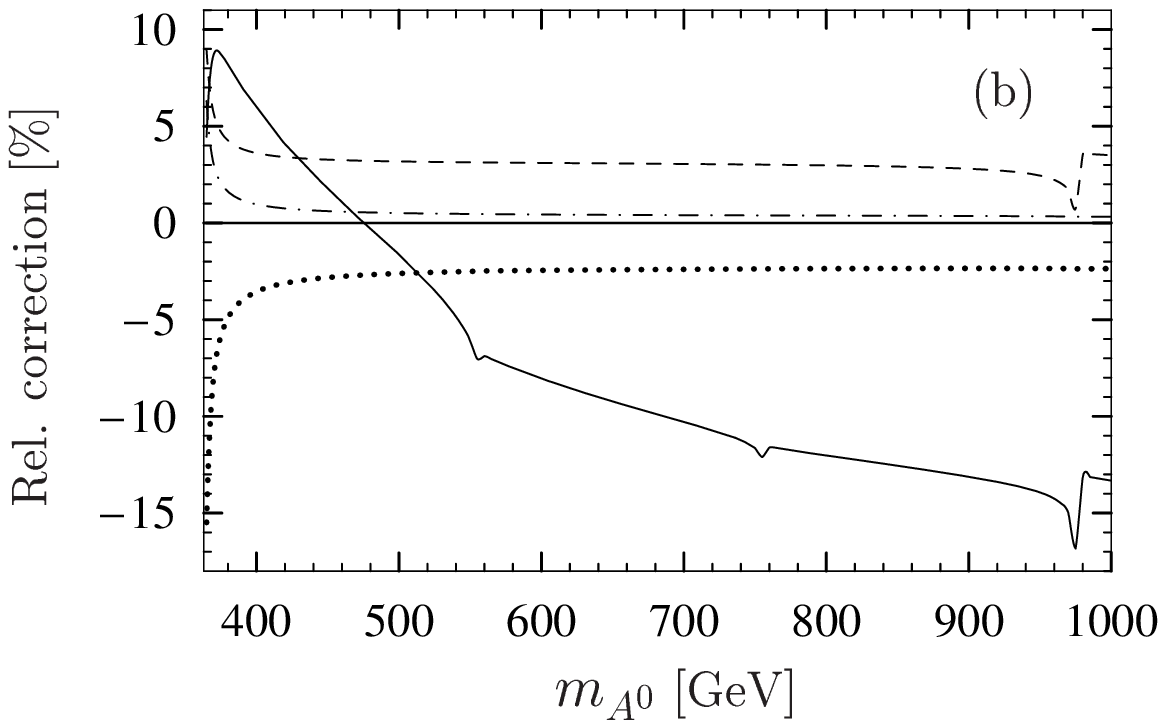}}}}}
\hspace*{-3mm}
\vspace{-2mm}
 \caption[fig2]
 {Naive tree-level (dotted), tree-level (dashed) and 
  one-loop corrected (solid) widths of 
  the decays $A^0\to\ch^+_1+\ch^-_1$ as functions of $m_{A^0}$. 
 The individual loop contributions to (a) in terms of the naive tree-level
width are shown in (b),
(s)fermion loop (dash-dotted) and full one-loop (dotted) $\ch$ mass matrix contr.,
total ($\ch$ mass matrix and naive corrections together) (s)fermion loop 
(dashed) and full (solid) one-loop contr.}
\label{fig:1}
 \end{center}
\end{figure}

Fig.~\ref{fig:1}~(a) shows the tree-level and corrected width of the decay
$A^0\to\ch^+_1+\ch^-_1$ as a function of $m_{A^0}$. The full one-loop correction goes up to
\mbox{$\sim -15$\%}. In Fig.~\ref{fig:1}~(b) the individual contributions in terms of the naive 
tree-level are given. This figure shows that all four contributions are of comparable order,
both for the $\ch$ mass matrix and the naive vertex correction.

\begin{figure}[h!]
 \begin{center}
\hspace*{-3mm}
\mbox{
\mbox{\resizebox{59mm}{!}{\includegraphics{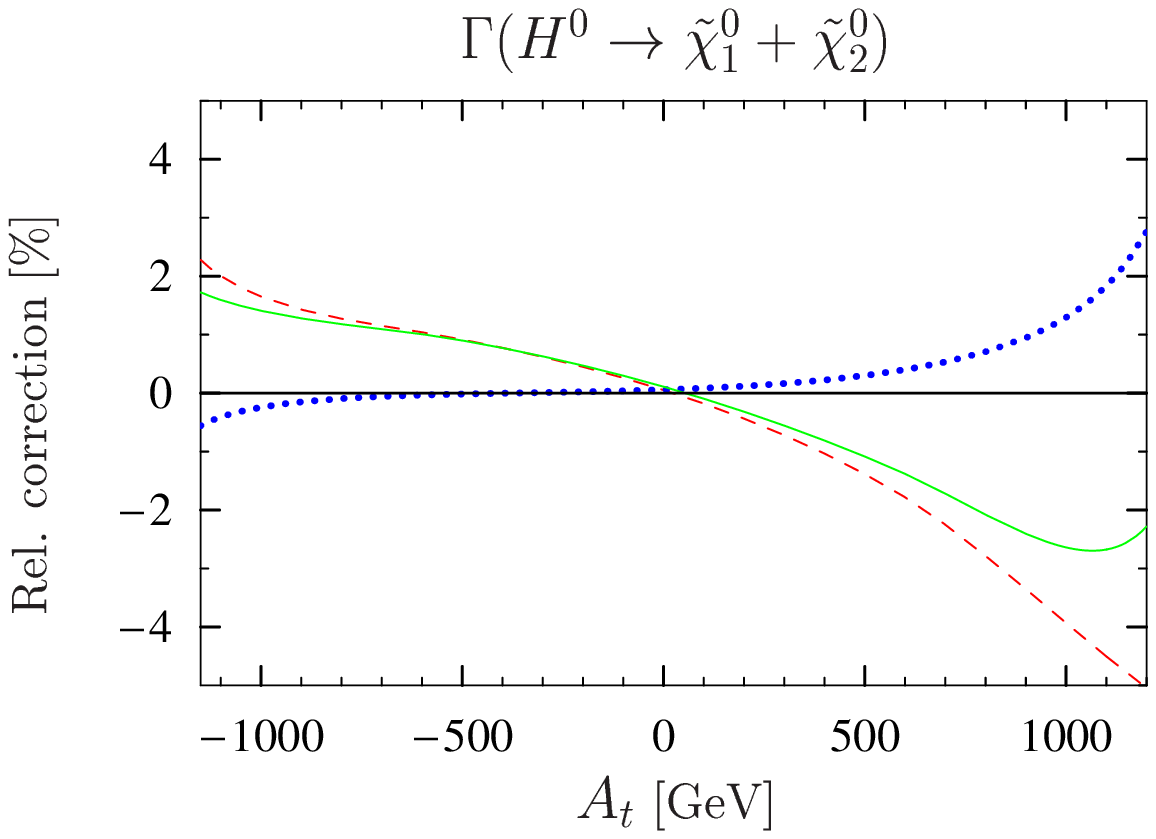}}} 
\raisebox{0.0mm}{\mbox{\resizebox{63mm}{!}{\includegraphics{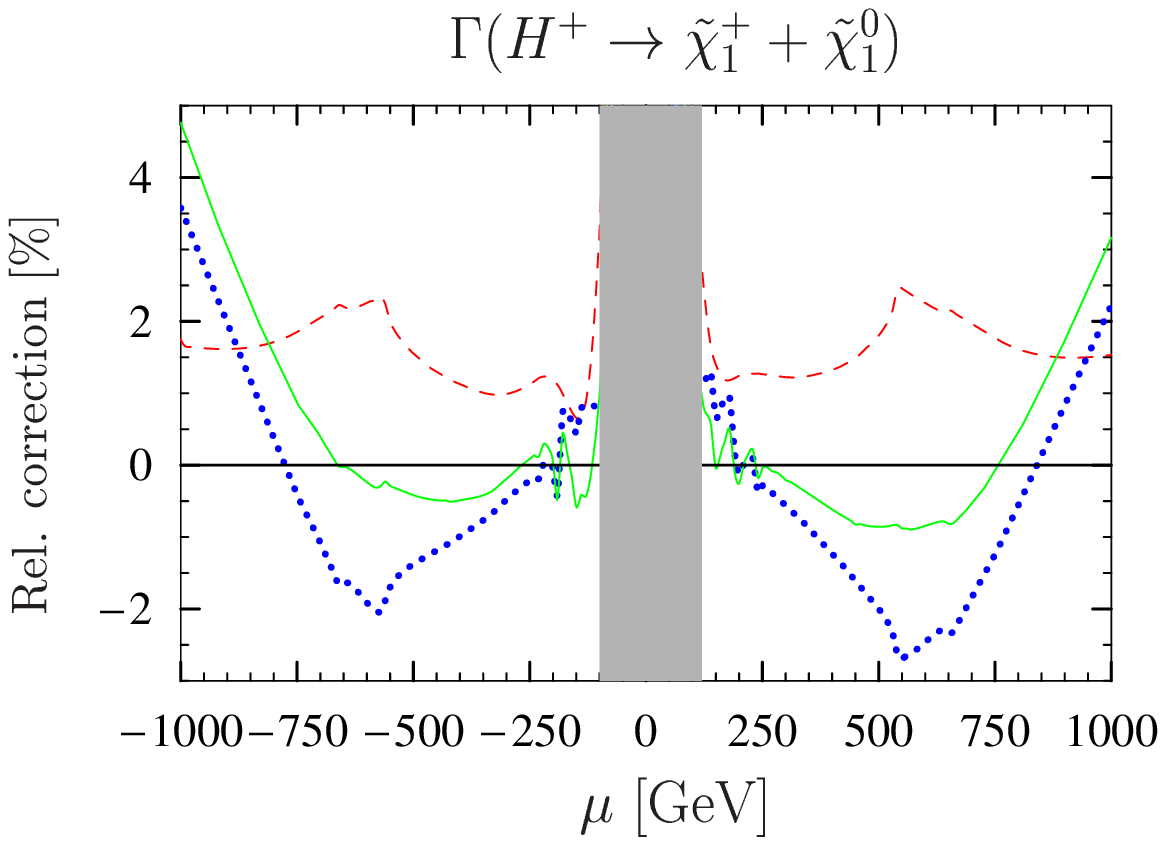}}}}}
\hspace*{-3mm}
\vspace*{-6mm}
 \caption[fig2]{Shift due to $\ch$-mass matrix renormalization (red-dashed), 
naive vertex (blue-dotted), and full one-loop (green-solid) correction, 
in terms of the naive tree-level
width shown for two decays.}  
\label{fig:2}
\end{center}
\end{figure}

Fig.~\ref{fig:2}~ show the decay $H^0\to\ch^0_1+\ch^0_2$ (left) as a function of 
the $A_t$ ($= A_b = A_\tau$) parameter and the decay
$H^+\to\ch^0_1+\ch^+_1$ (right) as a function of $\mu$.
The grey area is excluded by the $m_{\ch^+_1}$ LEP2 bound. The corrections can 
go up to $\pm 4$\%. Again, the $\ch$ mass matrix and the naive vertex correction
are of comparable size. In the right figure a few pseudo thresholds are seen due to 
opening channels.  Also the crossed channels, e.g. 
$\ch^0_3 \to h^0 + \ch^0_1$, are studied numerically. Due to lack of space these
figures will be presented elsewhere. 

\section{Conclusion}

We have calculated the full one-loop corrections to
the Higgs boson decays into charginos and neutralinos. 
All parameters in the chargino and neutralino mass matrix  and mixing matrices 
($U$, $V$, and $Z$) are renormalized in the on-shell scheme. 
We showed that the the corrections due to this renormalization are of 
similar size as the naive one-loop corrections.
We have studied the dependence of the corrections 
on the SUSY parameters. The corrections are typically of the order
of $\sim 10$\%. 
The (s)fermionic corrections were 
shown to be of similar order of magnitude. 
Therefore, for precision measurements the full one-loop corrections
due to the chargino and neutralino mass matrix renormalization and the naive
corrections have to be taken into account.


\section*{References}
\begin{thebibliography}{99}
 
\bibitem{chmasscorr1}
H. Eberl, M. Kincel, W. Majerotto, and Y. Yamada, 
\PRD 64 (2001) 115013;
W. \"Oller, H. Eberl, W. Majerotto, and C. Weber, 
\EPJ 29 (2003) 563. 

\bibitem{chmasscorr2}
T. Fritzsche and W. Hollik, \EPJ 24 (2002) 619. 

\bibitem{Higgs->charginos}
H. Eberl, W. Majerotto, and Y. Yamada, \PLB 597 (2004) 275, and 
references therein.

\bibitem{H0->neutralinos}
H. Eberl, M. Kincel, W. Majerotto, and Y. Yamada, 
\NPB 625 (2002) 372.

\bibitem{zhang} 
Zhang Ren-You, Ma Wen-Gan, Wan Lang-Hui, and Jiang Yi,
\PRD 65 (2002) 075018.  

\bibitem{tree2}
A. Djouadi, J. Kalinowski, and P. M. Zerwas,
{\em Z. Phys.} C 57 (1993) 569;
A. Djouadi, P. Janot, J. Kalinowski, and P. M. Zerwas,
\PLB 376 (1996) 220; 
A. Djouadi, J. Kalinowski, P. Ohmann, and P. M. Zerwas,
{\em Z. Phys.} C 74 (1997) 93. 

\bibitem{FeynArts}
T. Hahn, {\em Nucl. Phys. Proc. Suppl.} B 89 (2000) 231; 
{\em Comp. Phys. Commun.} 140 (2001) 418; 
{\it FeynArts User's Guide, FormCalc User's Guide} and 
{\it LoopTools User's Guide}, available at http://www.feynarts.de.  

\bibitem{SPS1a}
B. C. Allanach et al., \EPJ 25 (2002) 113. 

\end{thebibliography}
\end{document}